\documentclass[12pt,preprint]{aastex}
\usepackage{amsbsy}
\usepackage{amsmath}
\newcommand{\be}{\begin{equation}}
\newcommand{\ee}{\end{equation}}

\newcommand{\vv}{{\bf v}}

\newcommand{\pa}{\partial}

\newcommand{\Rey}{{\it Re}}
\newcommand{\f}{\frac}

\def\ltsima{$\; \buildrel < \over \sim \;$}

\def\lsim{\lower.5ex\hbox{\ltsima}}
\def\gtsima{$\; \buildrel > \over \sim \;$}
\def\gsim{\lower.5ex\hbox{\gtsima}}

\begin{document}
\title{Gravitational radiation from nonaxisymmetric
spherical Couette flow in a neutron star}

\author{C. Peralta,\altaffilmark{1,}\altaffilmark{2} A. Melatos,\altaffilmark{1} M. Giacobello\altaffilmark{3} and A. Ooi\altaffilmark{3} }

\email{cperalta@physics.unimelb.edu.au}

\altaffiltext{1}{School of Physics, University of Melbourne,
Parkville, VIC 3010, Australia}

\altaffiltext{2}{Departamento de F\'{\i}sica, Escuela de Ciencias,
Universidad de Oriente, Cuman\'a, Venezuela}

\altaffiltext{3}{Department of Mechanical Engineering, University
of Melbourne, Parkville, VIC 3010, Australia}

\begin{abstract}
\noindent 
The gravitational wave signal generated by
global, nonaxisymmetric shear flows in a neutron star is calculated
numerically 
by integrating the incompressible 
Navier--Stokes equation in
a spherical, differentially rotating shell. 
At Reynolds numbers $\Rey \gsim 3 \times 10^{3}$,
the laminar Stokes flow is unstable and helical, oscillating
Taylor--G\"ortler vortices develop. The gravitational wave
strain generated by the resulting kinetic-energy
fluctuations 
is computed in both $+$ and $\times$ polarizations 
as a function of time. 
It is found that the signal-to-noise
ratio for a coherent, $10^{8}$-{\rm s} integration with LIGO II scales
as $ 6.5 (\Omega_*/10^{4} \, {\rm rad}\,{\rm s}^{-1})^{7/2}$ for
a star at $1$ {\rm kpc} with angular velocity $\Omega_*$.
This should be regarded as a 
lower limit: it excludes pressure fluctuations,
herringbone flows, Stuart vortices, and 
fully developed turbulence 
(for $\Rey \gsim 10^{6}$).

\end{abstract}

\keywords{gravitational waves --- hydrodynamics --- stars: neutron --- stars: rotation}

\section{Introduction}

Gravitational radiation from {\it linear}, global, fluid oscillations
in neutron stars
has been studied extensively, e.g. radiation-reaction-driven r-modes
\citep{andersson98,lu01},
and two-stream superfluid oscillations with entrainment  
\citep{a03,pca04}.
However, gravitational radiation is also emitted by
{\it nonlinear}, global, fluid oscillations.
Specifically, it is well known that a viscous Navier--Stokes fluid inside
a spherical, differentially
rotating shell --- spherical Couette flow (SCF) ---
undergoes sudden transitions between states with different
vortex topologies, which are
generally nonaxisymmetric, due to shear instabilities 
\citep{mt87a,mt87b,je00}.
Recent simulations of superfluid SCF,
based on the Hall--Vinen--Bekarevich--Khalatnikov
(HVBK) theory of a $^1S_0$ neutron superfluid, 
confirm
that such transitions to nonaxisymmetric flows also
occur in neutron stars \citep{pmgo05,pmgo06}.

In this Letter, we compute the gravitational wave signal generated
by nonaxisymmetric flows in young, rapidly rotating neutron stars.
There are at least two astrophysical scenarios in which these develop.
First, if the crust and core of the star are loosely coupled, as
indicated by observations of pulsar glitches \citep{sl96,lss00},
differential rotation builds up 
as the crust spins down electromagnetically, until the core undergoes
a transition to the Taylor--G\"ortler vortex state,
a helical, oscillating flow that is one of the last
stages in the laminar-turbulent transition in SCF
as the Reynolds number increases \citep{n83,nt88}.
Second, if the crust precesses, while
the core does not, the rotation axes of the crust
and core misalign, 
inducing a rich variety of flow patterns 
and instabilities, like time-dependent shear
waves, leading to fully developed turbulence
\citep{wv95}.
We treat the former scenario here and postpone
the latter to a separate paper.

In \S\ref{sec:motioneq}, an SCF model
of a differentially rotating neutron star is proposed.
In \S\ref{sec:3Dflow},
we explore the transitional dynamics
and topology of the flow.
In \S\ref{sec:detectability}, we calculate the
gravitational wave
signal and its frequency spectrum.

\section{Global SCF model and numerical method}
\label{sec:motioneq}
We consider an idealized, two-component model of a neutron star, in which
a solid crust is loosely coupled
to a differentially rotating fluid core. Specifically,
we consider the motion of a neutron fluid within
a spherical, differentially rotating shell in the outer
core of the star, where the density $\rho$ lies in
the range $ 0.6 < \rho/\rho_c  < 1.5$, $\rho_c = 
2.6 \times 10^{14} \, {\rm g \, cm}^{-3}$ \citep{ss95}.
The inner (radius $R_1$) and
outer (radius $R_2$) boundaries
rotate at angular frequencies $\Omega_1$ and $\Omega_2$,
respectively, about the $z$ axis. As the star is strongly
stratified, the shell is thin, with $R_2 - R_1 \leq 0.1 R_1$
\citep{ae96,ld04}. The rotational shear is sustained by the
vacuum-dipole spin-down torque, violent events
at birth \citep{dfm02b,oblw04}, accretion in a
binary \citep{fujimoto93}, neutron star mergers \citep{su00}, or 
internal oscillations like r-modes \citep{rls00,lu01}.

We describe the fluid by
the isothermal Navier--Stokes equation, which, in
the inertial frame of an external observer, takes
the form 
\begin{equation}
\f{\pa \vv}{\pa t} + (\vv \! \cdot \! \nabla) \vv=
-\frac{\nabla p}{\rho} + \nu \nabla^2 \vv
+ \nabla \Phi
\label{eq:navier}
\end{equation}
in the incompressible limit $\nabla \cdot \vv =0$,
where $\vv$ is the fluid velocity,
$\nu$ is the kinematic viscosity,
$\rho$ is the density, $p$ is the pressure, and
$\Phi$ is the Newtonian gravitational potential. Henceforth,
$\Phi$ is absorbed into $p$ by replacing $p - \rho \Phi$ with $p$,
such that pressure balances gravity in the stationary
equilibrium ($\vv=0$) and there are no gravitational forces
driving the flow in the incompressible limit when the
stationary equilibrium is perturbed.
The Reynolds number and dimensionless gap width are defined
by $\Rey = \Omega_1 R_1^2/\nu$ and $\delta = (R_2 - R_1)/R_1$;
the rotational shear is $\Delta \Omega = \Omega_2-\Omega_1$. Equation
(\ref{eq:navier}) is solved subject to no-slip
boundary conditions.

Superfluidity plays a central role in the thermal
\citep{yls99} and hydrodynamic \citep{alpar78,r93,pmgo05}
behavior of neutron star interiors. However, 
we assume a viscous fluid here. Counterintuitively,
this is a good approximation because $\Rey$ ($\sim 10^{11}$)
is very high \citep{mm05}. It 
is known from terrestrial experiments on
superfluid $^4$He that, at high $\Rey$, quantized
vortices tend to lock the superfluid to turbulent
eddies in the normal fluid by mutual friction
\citep{bsbd97}, so that
the superfluid resembles classical Navier--Stokes
turbulence.
Moreover, in HVBK simulations of SCF,
it is 
observed that the global circulation pattern of
the superfluid does not differ much from 
a Navier--Stokes fluid at $\Rey \gsim 250$ \citep{hb04,pmgo05}.
Finally, a
viscous interior is expected in
newly born neutron stars, which are hotter than
the superfluidity transition temperature \citep{aks99}.

By assuming a viscous fluid, we omit from (\ref{eq:navier})
the coupling between quantized vortices 
and the normal fluid \citep{hv56a},
entrainment of superfluid
neutrons by protons \citep{c02},
and pinning of quantized vortices in the inner crust,
at radius $R_1$ [\citet{bel92}; cf. \citet{j98}].
We also neglect vortex pinning in the outer core
for simplicity, even though evidence exists that it may
be important when magnetic fields are included. Specifically,
three-fluid models of post-glitch relaxation based on the core dynamics
favor vortex pinning at the phase separation boundary of the core
\citep{ss95}, e.g. due to the interaction between vortex clusters
and the Meissner supercurrent set up by the crustal magnetic field
at the phase boundary \citep{sc99}. Other models analyze 
interpinning of proton and neutron vortices in the core
\citep{ruderman91c} and its effect on precession \citep{l03}. 
In the context of the hydrodynamic model in
this paper, pinning effectively increases the viscosity
of the core, reducing $\Rey$.

We use a pseudospectral collocation method to solve equation
(\ref{eq:navier}) \citep{bb02,giacobello05}.
The equations are spatially discretized in spherical
polar coordinates $(r,\theta,\phi)$, using 
restricted Fourier expansions in $\theta$ and $\phi$ 
and a Chebyshev expansion in $r$. The solution
is advanced using a two-step 
fractional-step method, which is accurate to second
order in the time-step \citep{canuto88}.
We limit ourselves to narrow gaps $\delta \leq 0.1$,
which are computationally less expensive and for which
more experimental
studies are available for comparison \citep{ybm77,n83}.
Recently, we performed the first stable simulations of 
superfluid SCF using this method \citep{pmgo05}.

\section{Nonaxisymmetric spherical Couette flow}
\label{sec:3Dflow}
The bifurcations and instabilities leading to transitions between SCF
states are controlled by three parameters:
$\delta$, $\Rey$, 
and $\Delta \Omega$.
Additionally, the history of the flow
influences its post-instability evolution
and the final transition to turbulence
\citep{wer99,je00}. For example, in experiments
on Taylor-G\"ortler vortices (TGV), the final state depends on 
$d\Omega_1/dt$, i.e. on the time to reach the critical $\Rey$ relative to the viscous
diffusion time \citep{nt95}.
In general, four SCF states can be distinguished: (i) a laminar
basic flow, (ii) a toroidal or helical Taylor--G\"ortler
flow, (iii) a transitional flow with nonaxisymmetric,
oscillating TGV, and (iv) fully developed turbulence
\citep{ybm75,n83}. 
In a neutron star, where
$\Rey \sim 10^{11}$, the state is probably
turbulent. However,
we are restricted by our computational resources
to simulate regimes (i) and (ii),
with $1 \times 10^3 \leq \Rey \leq 3 \times 10^{4}$.

\subsection{TGV transition: initial conditions}
\label{sec:tgv}
Experimentally, for $\delta =0.06$, TGV are
obtained by quasistatically increasing $\Omega_1$
from zero to give $\Rey \sim 3300$  \citep{nt05}.
This procedure can be painfully slow, because a steady
state must be reached at each intermediate step.
Numerically, we circumvent it
by following \citet{li04} and introducing a nonaxisymmetric perturbation 
with azimuthal wave number $m_a$ of the form 
\begin{eqnarray}
\label{eq:perturb}
v_r & = &  - \frac{4 \epsilon_1 (r-R_1)(r-R_2)}{\delta^2 R_1^2} 
\cos \chi (\theta, \phi), \\
\label{eq:perturb2}
v_\theta & = &  \frac{\epsilon_1}{\delta} \sin\left[
\frac{\pi (r-R_1)(r-R_2)}{2 (R_2 -R_1)^2}\right]
\sin \chi (\theta, \phi),
\end{eqnarray}
with $\chi = \pi[1-R_2(\pi/2-\theta)/(R_2 -R_1)-0.4 \sin(m_a \phi)]$.
This is done as follows. Before introducing the perturbation, a steady
state for $\Rey=2667$ is obtained (just below the critical $\Rey$
where TGV emerge experimentally). Then, the Reynolds number is
raised instantaneously to $\Rey=3300$, and we continue by adding
(\ref{eq:perturb}) and (\ref{eq:perturb2}), with
amplitude $\epsilon_1 \sim 10^{-6} \, \Omega_1 R_1$, 
to the numerical solution at each time-step until
a viscous diffusion time $t_d = (R_2 - R_1)^2/\nu$ elapses.
We then stop adding the perturbation
and the flow is left to evolve according to (\ref{eq:navier}),
until a final steady state is reached. More than one
perturbation can lead to the same final state; some authors
add Gaussian noise \citep{z96}, although this
affords less control over the wavenumber and the number of
vortices excited.

The perturbation excites TGV by shedding 
vorticity from the inner sphere \citep{li04}.
Perturbations with $2 \leq m_a \leq 5$ are explored,
supplementing experiments with $m_a=3$ \citep{n83}.
For numerical simplicity, we limit
ourselves to the case where only the inner sphere rotates.
TGV with both spheres
rotating are equally possible and have been observed experimentally; they
exhibit additional twisting (and increased nonaxisymmetry) in the 
helical vortices \citep{nt05}, so the gravitational wave strain we
compute in \S\ref{sec:detectability}
is a lower limit.

\subsection{Flow topology}
\label{sec:topology}

Figure \ref{fig:fig1}a shows a
kinetic-energy-density isosurface $(0.11 \rho R_1^5 \Omega_1^2)$ for 
the fully developed
TGV state $m_a =3$, $\delta=0.06$, $\Rey=3300$. Its nonaxisymmetry
is apparent in 
the striated equatorial
bands, whose inclination with respect to the equator varies
with longitude from zero to $\sim 3$ {\rm deg} \citep{n83}.

A popular way to classify complex, three-dimensional
flows is to construct scalar invariants 
from $A_{ij}=\partial v_i/\partial x_j$ 
\citep{cpc90,jh95}. Specifically,
the discriminant
$D_A = Q_A^3 + 27 R_A^2/4$, with $R_A = -{\rm det}(A_{ij})$
and $Q_A = (A_{ii}^2-A_{ij} A_{ji})/2$,
distinguishes between
regions which are focal ($D_A > 0$) and strain-dominated
($D_A < 0$). Figure \ref{fig:fig1}b plots the $D_A = 10^{-3} \, \Omega_1^{6}$
isosurface in color: yellow regions are
stable focus/stretching, \footnote{Trajectories are repelled away
from a fixed point along a real eigenvector of $A_{ij}$ (stretch) and
describe an inward spiral when projected onto the plane normal
to the eigenvector (stable focus).} blue regions are unstable focus/contracting. 
The filaments coincide with the TGV.
One circumferential vortex and $m_a=3$ helical
vortices lie in each hemisphere.
We also plot the isosurface $D_A = -10^{-3} \, \Omega_1^{6}$:
red and green regions have stable node/saddle/saddle 
and unstable node/saddle/saddle topologies respectively
\citep{cpc90}.
The helical vortices span
$\sim 40$ {\rm deg} of latitude 
and travel in the $\phi$ direction, with
phase speed $\approx 0.48 \Omega_1 R_1$.

The TGV state is oscillatory, which is important
for the
gravitational wave spectrum.
Its periodicity is evident in Figure \ref{fig:fig1}c, where
$v_\theta$ at an equatorial point is plotted
versus time for $2 \leq m_a \leq 5$.
For $m_a=3$, $v_\theta$ oscillates with
period $4.4 \Omega_1^{-1}$, the time for
successive helical vortices to pass by a stationary observer, in 
accord with
experiments \citep{n83}.
Instantaneous streamlines are drawn in three meridional
planes in Figure \ref{fig:fig1}d, highlighting the 
nonaxisymmetry.
One obtains a similar TGV state for $\delta=0.14$, $=0.18$ \citep{sn01,li04}.

\section{Gravitational wave signal}
\label{sec:detectability}

The metric
perturbation $h_{ij}^{\rm TT} = (4 G/r c^4) \int d^3 {\bf x}\, T_{ij}$ 
in the transverse-traceless gauge
can be calculated using Einstein's quadrupole
formula \citep{gravitation}
\begin{equation}
\label{eq:formulita}
h_{ij}^{\rm TT} = \frac{2 G \rho}{r c^{6}} \frac{\partial^2 }{\partial t^2} \int d^3 {\bf x} \, |\vv|^2
\left(x_i x_j - \delta_{ij} \frac{|{\bf x}|^2}{3}\right),
\end{equation}
where the integral is over the source volume, $r$ is the distance to the source, and
$T_{ij} = \rho \gamma^2 v_i v_j + p \delta_{ij}$ is the
stress-energy of
a Newtonian fluid.
Note that we approximate $T_{00} = \rho \gamma^2 \approx \rho (1+|\vv|^2/c^2)$
in (\ref{eq:formulita}), and we omit
a thermal-energy contribution $\propto \delta p$ ($\sim \rho |\vv|^2$,
by Bernoulli's theorem), which we cannot calculate with our incompressible
solver. 
Far from the source, we write 
$h^{\rm TT}_{ij} = h_+ e_{ij}^+  + h_\times e_{ij}^\times$, with the 
polarizations defined
by $e_{yy}^+ = -e_{zz}^+=e^{\times}_{yz} = e^{\times}_{zy}=1$
for an observer on the $x$ axis.

\begin{figure*}
\epsscale{0.7}
\plotone{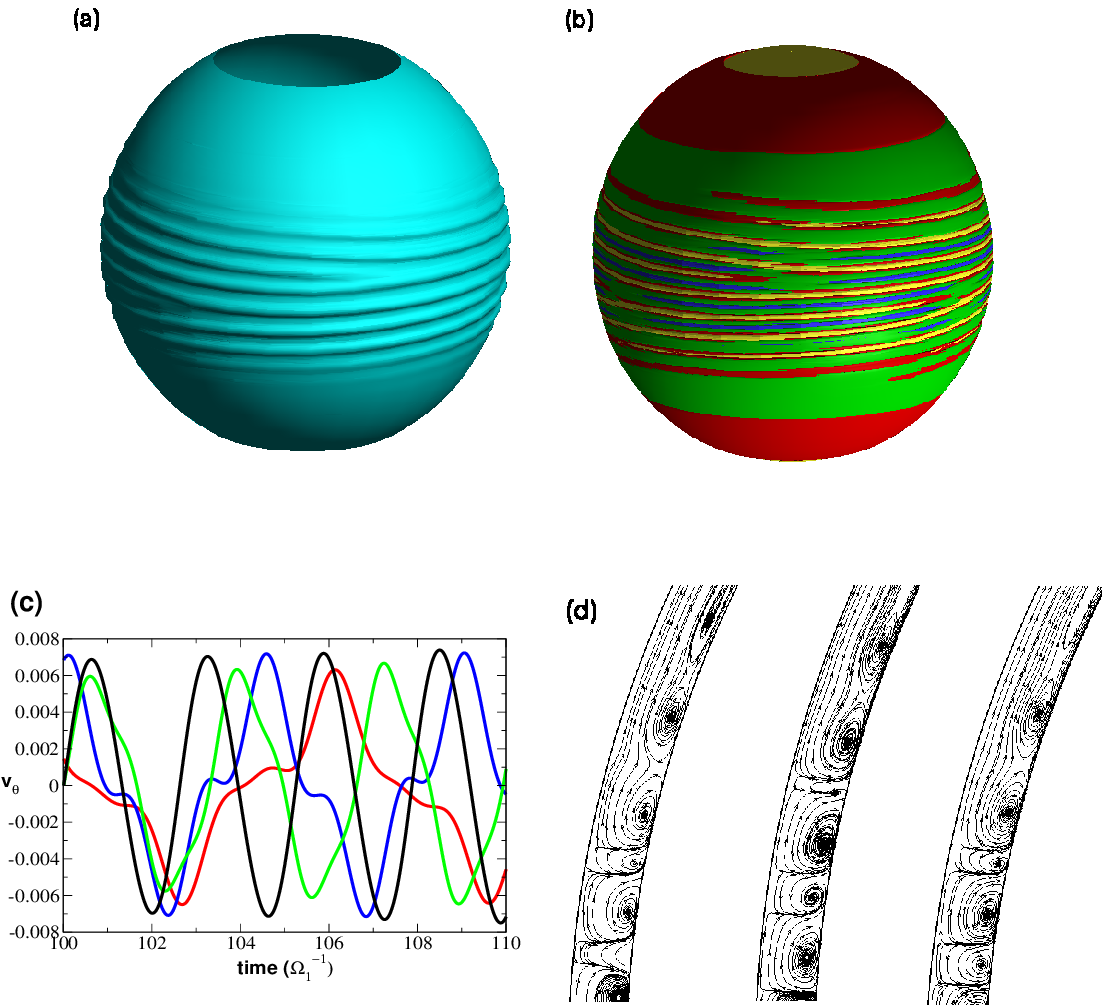}
\caption{Taylor-G\"ortler vortices for 
$m_a=3$, $\delta=0.06$, at $t=100 \, \Omega_1^{-1}$.
(a) Isosurface of
kinetic energy: $\rho |\vv|^2=0.11 \rho R_1^5 \Omega_1^2$. 
(b) Isosurfaces of velocity-gradient discriminant: $D_A = 10^{-3} \, \Omega_1^6$,
stable focus/stretching
(yellow) and unstable focus/contracting (blue);
$D_A = -10^{-3} \, \Omega_1^6$,
stable node/saddle/saddle (red) and
unstable node/saddle/saddle (green). 
(c) Time history of $v_\theta$ at $r=1.06$, $\theta=\pi/2$, $\phi=0$
for $m_a=2$ (red), $3$ (blue), $4$ (green), and $5$ (black).
(d) Streamlines obtained by integrating
the in-plane components of $\vv$ in the planes
$\phi=0$ (left), $\phi=\pi/2$ (middle), and $\phi = 2\pi/9$ (right).}
\label{fig:fig1}
\end{figure*}

\begin{figure*}
\epsscale{0.7}
\plotone{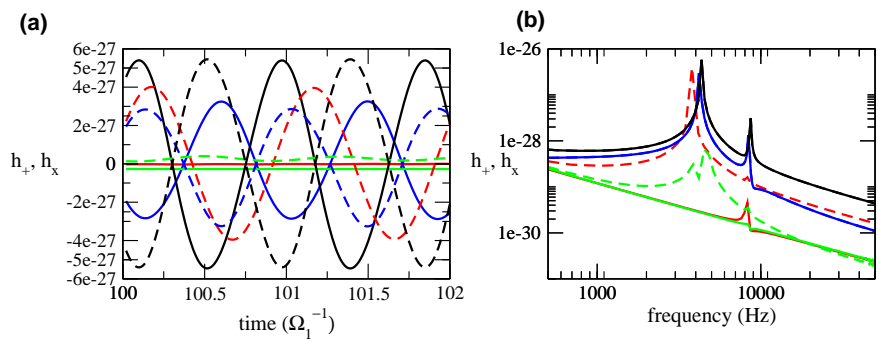}
\caption{Gravitational waves from Taylor-G\"ortler vortices in
a neutron star with $\Omega_1/2\pi=600$ {\rm Hz} and $r=1$ {\rm kpc}.
(a) Signal $h_{+,\times} (t)$, and (b) frequency spectrum $h_{+,\times} (f)$, for
$m_a=2$ (red), $3$ (blue), $4$ (green), and 
$5$ (black). Solid and dashed curves correspond to the $+$ and $\times$
polarizations respectively, as measured by an observer on the $x$ axis.}
\label{fig:fig2}
\end{figure*}

In Figure \ref{fig:fig2}a, we plot the $h_+$ and $h_\times$ 
polarizations versus time, as seen by an observer
on the $x$ axis, for $2 \leq m_a\leq 5$ (the signal
is $\sim 3$ times weaker on the $z$
axis, where the north-south asymmetry is not seen in projection).
The amplitude is greatest for $m_a = 5$ but depends weakly
on $m_a$. {\it Importantly, $h_+$ and $h_\times$ are
$\pi/2$ out of phase for all $m_a$, and we find $|h_+| \ll |h_\times|$
for even $m_a$}. These two signatures  offer a promising target
for future observations.
The period ($\approx 0.8 \Omega_1^{-1}$) is similar for
$3 \leq m_a \leq 5$, where a maximum of three helical vortices
are excited \citep{li04} (we
find that the torque at $R_2$ oscillates with the same period).
This too is good for detection, because
several modes are likely to be excited in a real star. The period
arises because the isosurface in Figure \ref{fig:fig1}a forms
a pattern with six-fold symmetry when projected onto the $y$-$z$
plane, with fundamental period $4.4 \Omega_1^{-1}$ 

In Figure \ref{fig:fig2}b, we present the frequency spectra
of the $+$ and $\times$ polarizations. 
The two peaks,
at $f \approx 1.2 \Omega_1$ and  $2.4 \Omega_1$, have
full-width-half-maxima of $\sim 300$ {\rm Hz}. This
is caused by the subharmonics evident in Figure \ref{fig:fig1}c, which
arise because the northern and southern helical vortices start
at unequal and variable longitudes. {\it In addition,
the peaks for even and odd $m_a$ are displaced by
$\sim 500$ {\rm Hz}, and the primary peak for $m_a=4$
is split}. These two spectral signatures are a promising
target for future observations. The even-odd displacement may
arise because the phase speeds of the vortex patterns
for $m_a=2$, $3$ differ by $\approx 0.01 \Omega_1 R_1$ \citep{li04}.
For large $f$, we find $h_{+, \times}(f) \propto f^{-3/2}$.

The squared signal-to-noise ratio can be calculated from 
$(S/N)^2 = 4 \int_0^\infty df |h(f)|^2/S_h(f)$ \citep{creighton03},
where we take $S_h(f) = 10^{-50} (f/0.6 \, {\rm kHz})^2$ 
($ 0.2 \leq f \leq 3$ {\rm kHz}) for a $10^8$-{\rm s} integration
with LIGO II,
if the frequency and phase of the signal  
are known in advance \citep{bccs98}.
For a star with 
$\Omega_1/2\pi=600$ {\rm Hz}, at $r=1$ {\rm kpc}, we
find $S/N=0.21$ ($m_a=2$), 
$0.12$ ($m_a=3$), $0.023$ ($m_a=4$),
and  $0.22$ ($m_a=5$), although
there will be some
leakage of signal when the coherent integration is performed, because
the spectrum is not monochromatic.
For $m_a \leq 5$, $S/N$ increases (decreases) as $m_a$ odd (even)
increases.
We also find $S/N \propto \Omega_1^{7/2}$, which implies that
the fastest millisecond pulsars ($\Omega_1/2\pi \sim 1$ {\rm kHz})
and newly born pulsars with $\Delta \Omega / \Omega_1 \sim 0.1$ are
most likely to be detected.
The flow states leading to turbulence at $\Rey \gsim 10^6$, e.g.
shear waves,
Stuart vortices, and ``herringbone" waves \citep{nt88},
make contributions
of similar order to the TGV signal in Figure \ref{fig:fig2}.

The Reynolds number in the outer core of a neutron star,
$\Rey = 10^{10} (\rho/10^{15} \,{\rm g}\,{\rm cm}^{-3})^{-1}$
$( T/10^8 \, {\rm K})^2 (\Omega_*/10^4\,{\rm rad}\,{\rm s}^{-1})$
\citep{mm05}, where $T$ is the temperature,
typically exceeds the maximum 
$\Rey$
in our simulations (due to computational capacity). For
$\Rey \gsim 10^6$, the flow
is turbulent and therefore
axisymmetric when time-averaged --- but not instantaneously.
In Kolmogorov turbulence, the characteristic flow speed in an
eddy of wavenumber $k$ scales as $k^{-1/3}$, the turbulent
kinetic energy scales as $k^{-5/3}$, and the
turn-over time scales as $k^{-2/3}$. From equation (\ref{eq:formulita}), 
the {\it rms} wave strain scales as 
$k^{-3/2} \int dk \, k^{-1/3}$ after averaging
over Gaussian fluctuations, and
is dominated by the largest eddies (where most of the kinetic energy
also resides).
We therefore
conclude that, even for $\Rey \gsim 10^{6}$,
the largest eddies, which resemble organized
structures like TGV, dominate $h_+$ and $h_\times$, not 
the isotropic small eddies, but the signal is noisier.
Hence, counterintuitively, we predict that hotter neutron stars
(with lower $\Rey$) have narrower gravitational wave spectra than
cooler neutron stars, {\it ceteris paribus}.

We ignore compressibility when computing
$T_{00}$ ($\delta p =0$) and
solving (\ref{eq:navier}) numerically
by pressure projection.
Yet, realistically,
neutron stars are strongly stratified \citep{ae96}, confining
the meridional flow in Figure \ref{fig:fig1}d into narrow
layers near $r=R_1$, $R_2$. Stratification
can be implemented crudely by using
a low-pass spectral filter to artificially suppres $v_r$ \citep{don94,pmgo06}. 
We postpone this to future work.

\acknowledgments We acknowledge
the computer time supplied by the Australian Partnership for
Advanced Computation (APAC). We thank Prof. Li Yuan,
from the Chinese Academy of Sciences, for very helpful 
discussions.


\newpage

\end{document}